\input{psfig}
\psfigurepath{./fig/}
\documentstyle [fleqn,12pt] {article}
\newcommand{\al}{\alpha}

\newcommand{\de}{\delta}
\newcommand{\De}{\Delta}
\newcommand{\ep}{\epsilon}
\newcommand{\ga}{\gamma}

\newcommand{\la}{\lambda}
\newcommand{\Om}{\Omega}

\newcommand{\Si}{\Sigma}
\newcommand{\th}{\theta}

\newcommand{\vi}{\varphi}

\newcommand{\be}{\begin{equation}}
\newcommand{\ee}{\end{equation}}
\newcommand{\bea}{\begin{eqnarray}}
\newcommand{\eea}{\end{eqnarray}}
\newcommand{\bean}{\begin{eqnarray*}}
\newcommand{\eean}{\end{eqnarray*}}

\newcommand{\pl}{\mbox{Planck}}

\begin{document}          

\begin{titlepage}

\vspace*{2 cm}

\begin{center}
{\huge The Oscillating Universe: \\  \vspace{3mm}
an Alternative to Inflation \vspace{1.5 cm}}

Ruth Durrer$^\sharp$ and Joachim Laukenmann$^\flat$ \vspace{1cm}

$^\sharp$ Universit\'e de Gen\`eve, D\'epartement de Physique 
Th\'eorique,\\
24, quai Ernest Ansermet, CH-1211 Gen\`eve, Switzerland \vspace{5mm}

$^\flat$  Universit\"at Z\"urich, Institut f\"ur Theoretische Physik,\\
Winterthurerstrasse 190, CH-8057 Z\"urich, Switzerland 
\vspace{2cm}\\
\end{center}
\begin{abstract}
The aim of this paper is to show, that the 'oscillating universe' is a viable
alternative to inflation. We remind that  this model
provides a natural solution to the flatness or entropy and to the horizon
problem of standard cosmology. We study the evolution of density perturbations
and determine the power spectrum in a closed universe. The
results lead to constraints of how a previous cycle might have looked like.
We argue that most of the radiation entropy of the present universe may
have originated from gravitational entropy produced in a previous cycle.

We show that measurements of the power spectrum on very large scales could
in principle decide whether our universe is closed, flat or open.

\end{abstract}

\end{titlepage}

\section{Introduction}
 In a closed universe, the inevitable big crunch might actually be
followed by a subsequent big bang. This idea is very old. It goes back to
Lema\^{\i}tres Phoenix picture of 1933 \cite{Le}. As we shall remind
especially  younger readers, this 'oscillating universe'
provides quite naturally a solution to the flatness or entropy and to
the horizon problem. This was  well known before the advent
of inflationary models around 1980 in seminal papers by Starobinski, 
Guth, Linde and others \cite{inf}. Since then, it has been so 
completely forgotten, that
we ourselves originally believed to be the first to
study these ideas and older colleagues had to refer us to the original
literature. However, it is not only inflation which let people forget
about the oscillatory universe, but also an argument in essence due to 
Penrose \cite{Pe} that black hole formation in a previous cycle leads to 
far too much entropy and thereby to an even more severe entropy problem
in the opposite sense than the usual one.

It is the aim of this paper to show that  under most realistic
circumstances Penrose's conclusion need not be drawn.
We shall see that some amount of entropy production due to gravitational
clumping can just somewhat accelerate the growth of the maximal scale
factor, $a_{\max}$, from one cycle to the next, without over producing entropy.
This will lead us to the conclusion that the oscillating universe remains
a viable alternative to inflation. 

We consider this especially important since inflation has become some
kind of 'cosmological dogma' during the last ten years, despite the
fact that no inflationary scenario which solves the horizon and
flatness problems and yields acceptably small density fluctuations
has yet been constructed without  substantial  fine tuning (which may
 be protected by a symmetry and thus be 'technically natural').
Furthermore, many inflationary models are built upon the gravitational
action of vacuum energy, which acts gravitationally like a cosmological 
constant, the most miraculous number in cosmology,
which today is by a factor of about $10^{100}$ times smaller than what
we would expect from particle physics \cite{We}. This unbelievable amount
of fine tuning tells us, that our understanding of the interplay of gravity
and quantum vacuum energy is unsufficient. Therefore, a mechanism relying 
mainly on this interplay, to us, seems unsatisfactory.

Besides solving the horizon and entropy problems, inflation generically
predicts a scale invariant Harrison--Zel'dovich initial spectrum of 
fluctuations as it was observed by the DMR experiment on the COBE satellite
\cite{COBE}. These observations have been considered as great success of 
even 'proof' of inflation. However, also global topological defects \cite{DZ}
or cosmic strings \cite{HK}, which can form during phase transitions in the 
early universe,
naturally lead to a scale invariant spectrum of  fluctuations but they cannot 
easily be reconciled with inflation. 

 These considerations prompted us to look for possibilities to solve
the flatness and the horizon problem without invoking an inflationary
period. 

The basic picture which we  work out in this paper is the following:\\
The first 'big bang', the first 3--dimensional closed universe,  emerged
from quantum fluctuation of some, e.g., string vacuum. Its duration
was of the order of a Planck time. Due to some non thermal processes
there may have been a small gain of entropy, 
$S_{in}^{(1)}\le S_{end}^{(1)}$. The
first big crunch triggered the formation of  the next big bang who's
entropy was slightly larger and therefore its duration was slightly
longer, $S_{in}^{(1)}\le S_{end}^{(1)}\le S_{in}^{(2)}$.
This process continues with ever longer cycles. In a cycle with duration
significantly longer than Planck time, we assume, that
after a few Planck times during which the universe may have been in
some quantum gravity or stringy state, we have a mainly classical, 
radiation dominated universe. With the exception of  short periods
during which matter and radiation fall briefly out of thermal equilibrium,
the universe expands and contracts adiabatically. 

Since we have no theory
for the string or quantum gravity era of cosmology, it is very difficult to
estimate the number of Planckian cycles before the universe enters a
radiation dominated era. Furthermore, the entropy producing processes 
at high temperature depend on the microphysics at very high energy. We thus do
not have the tools to estimate the number of cycles which never entered 
a low temperature era. But, as we shall see, due to the production of 
gravitational  entropy, the number of cycles which entered a matter 
dominated era is very limited. 

As long as the universe remains radiation dominated no black holes can 
form. Only during a cycle
with a long enough  matter dominated era black hole
formation is possible. Penrose's argument now goes as follows: During
 a matter dominated era (small) black holes can form. These
finally, during the collapse phase, coalesce into one huge black
hole which eventually  contains the whole mass of the universe. Its
entropy is thus given by 
\be
	S_{bh}=(1/2)A_{bh}/G =2\pi R_s^2/G= 8\pi GM_{bh}^2\ge 10^{124},
	\label{ebh}
\ee
where we have set $M_{bh}$ equal to the present  mass within one 
Hubble volume which is of order $10^{23}M_\odot$ to obtain the last 
inequality. Clearly, already a significantly smaller mass would do, 
since the actual radiation entropy within the present Hubble
radius is of the order of $S_{Hubble}\sim 10^{87}$, a discrepancy of 
nearly 40 orders of magnitude. In terms of entropy per baryon, this yields
$\eta^{-1}= S/N_B \approx 10^{44}$ instead of the observed value 
$\eta^{-1}\approx 10^9$.

Is there a  way out of this simple but disastrous conclusion? \\
The first objection is that the laws of black hole
thermodynamics which rely heavily on Hawking radiation, hold only in
asymptotically flat spacetimes. Or, at least, that the entropy of a
black hole can be set equal to its area only for an observer outside
the black hole itself. Therefore, the black hole entropy formula
should only be adopted for black holes much smaller than the size of
the universe. Let us therefore add only the entropy of black holes
which are at least 10 times smaller than the curvature radius of the
universe and neglect subsequent growth of entropy due to the
coalescence of these black holes. Of course, this rule is somewhat ad
hoc, but as long as we have no clue of how to calculate in general the
entropy of the gravitational field it is a possible 'rule of 
thumb'. However, with this correction we gain only about a factor of
$10$ in the above entropy formula (\ref{ebh}) and not the required
factor of about $10^{38}$. But there is an additional short come in the
Penrose conclusion: The radiation entropy which we observe today is
the entropy generated mainly during the {\em previous} cycle whose 
matter dominated epoch might have been much shorter, leading to 
significantly less clumping and thus much less gravitational entropy 
production.

Furthermore, as we shall see, it is not clear that structure forms via
hierarchical clustering. In a pure radiation universe, it may well be
 that large black holes form first (if at all!) and the black hole 
entropy formula cannot be applied.

From these arguments it should be clear, that Penrose's objection
to the oscillating universe
does not have to be accepted and there may be ways out.
Another possibility not investigated in this work  is Israel's idea of
mass inflation inside the horizon of black holes \cite{Is,IS}.
Israel et al. accept the black hole entropy formula, but argue that 
inside the black hole horizon mass inflation takes place such that
the ratio $\eta^{-1}$ gets reduced substantially.

The reminder of this paper is organized as follows: In the next
section we give a brief review of the oscillatory universe. In the 
main part of this paper, section~3, we investigate cosmological 
perturbation theory in a closed universe and determine the evolution of 
a Harrison Zel'dovich initial spectrum in a purely radiation dominated 
universe and in a universe with an intermediate  matter dominated epoch.
We discuss how the final spectrum depends on the duration of the matter
dominated epoch and we formulate a limit for the maximum radiation
entropy of the previous cycle. The final section is devoted to our 
conclusions.

{\bf Notation:} The scale factor of the Friedmann universe is denoted by
$a$, we use the conformal time coordinate and the metric signature 
$(-,+,+,+)$, so that the Friedmann metric is given by
\be
	ds^2=a^2(dt^2-\ga_{ij}dx^idx^j)~,
\ee
where $\ga_{ij}$ is the metric of the unit three sphere, e.g.,
\[	\ga_{ij}dx^idx^j = d\chi^2+
	\sin^2(\chi)(d\th^2+\sin^2\th d\varphi^2)~.\]
Cosmic time is denoted by $\tau$, $\tau =\int^tadt$.\\
We normalize the scale factor $a$ such that the curvature of the spatial
sections is equal to $1/a^2$. 

\section{Reminder to the oscillatory universe}
Let us first explain how the oscillatory universe solves the flatness
or entropy problem. To do this it is useful to state the problem in
a somewhat different form: If the equation of state in a Friedmann
universe satisfies the strong energy condition, $\rho +3p>0$, then
$\rho$ decays faster than $1/a^2$ and $\Om=1$ is the unstable 'initial'
fix point of expansion. This means each Friedmann universe starts out
at $t\approx \mbox{ a few } t_{\mbox{Planck}}$ with $\Om\approx 1$ 
and later 
deviates more and more from this value. The flatness problem can thus
be stated as follows: How can it be, that our universe at its old age,
$t\gg t_{\mbox{Planck}}$, $T\ll T_{\pl}$ still looks so young, $\Om\sim 1$?
This problem is easily solved in the oscillating universe as we shall
now show.

The following arguments are due to Tolman \cite{To}. Only
 a year after Lema\^{\i}tre first brought up his phoenix picture Tolman
realized:
Since the entropy of the next
universal expansion can only be larger than the previous one, the
maximum expansion factor of the next cycle,
$a_{\max}$, is larger than the corresponding maximum of the previous
cycle. Since the density parameter $\Om$ starts deviating from 1
 only when the scale factor $a$ approaches $a_{\max}$, in the next
cycle it will take longer until this happens. We consider now
a cycle with a duration substantially longer than Planck time which
has entered a radiation dominated phase. If relativistic matter is 
in thermal equilibrium (which we assume to be true most of the time)
its energy density and entropy density are given by 
($\hbar=c=k_{\mbox{Boltzmann}}=1$)
\bea
	\rho &=& {\pi^2\over 30}NT^4	\label{rrad}\\
	s &=& {2\pi^2\over 45}NT^3 ~,	\label{srad}
\eea
where $N$ denotes the effective number of degrees of freedom (spin states).
$N= N_b +(7/8)N_f$. Here $N_b$ are bosonic degrees of freedom and $N_f$
are fermionic degrees of freedom.  Furthermore, from Friedmann's equation
for a closed universe,
\be \left({\dot{a}\over a}\right)^2 +1 
	= {8\pi G\over 3}a^2\rho ~,  \label{fried}\ee
together with (\ref{rrad}) and (\ref{srad}) one finds
\be 
	\Om-1={\rho-\rho_c\over\rho_c}=
	{8\pi G\rho a^2- 3(\dot{a}/a)^2\over 3(\dot{a}/a)^2}= {1 \over
   G\left({4N\pi\over 45}\right)^{1/3}S^{2/3}T^2 -1}
\label{om-1} ~.\ee
Here $S$ is the total entropy of the universe, $S=2\pi^2a^3s$.
Therefore, the larger the total entropy $S$ the
smaller the deviation of $\Om$ from the critical value 1 at a given
temperature $T$, or the lower temperatures are required for a substantial
increase of $\Om$. Expressing the maximal scale factor, $a_{\max}$, the
age of the universe at maximal expansion, $\tau_{\max}=\tau(a_{\max})$ and
the minimal temperature, $T_{\min}=T(a_{\max})$, in terms of the entropy
also show that these values grow, respectively decrease with
increasing entropy:
\bea
 a_{\max} &=& \la_1S^{2/3}~~,~~~ 
	\la_1=\left({45G^3\over 4\pi^7N}\right)^{1/6}
	 ~,\label{amax}\\
 \tau_{\max} &=& \int^{t_{\max}}adt = a_{\max} ~~~,~~~~ t_{\max}=\pi/2  ~,
	\label{tmax}\\
 T_{\min} &=&  \la_2S^{-1/3} ~~,~~~
     \la_2=\left({45\over 4\pi G^3N}\right)^{1/6}
	~.\label{Tmin}\eea
 The time it takes for the density parameter to differ
significantly from $1$ is a substantial fraction of $\tau_{\max}$.
Therefore, the
universe 'looks young' for longer and longer times as the entropy
increases  cycle by cycle.

It is clear that in the oscillating universe also the horizon problem
disappears since the age of the universe is not given approximately by
the inverse Hubble time, which is the age of the present cycle, but
the sum of the ages of all previous cycles has to be added, leading
to a much larger age which might even be infinite. For this solution
to be valid, it is important that correlations are not lost during a
big crunch/big bang passage and that the behavior of particles or
strings during this time is governed by a causal theory. A similar
problem is also encountered in 'pre--big--bang' string cosmology \cite{GV}.
 
In Appendix~B, we explore the possibility that quantum gravity 
may effectively lead to an Euclidean region of spacetime close to the
big crunch/big bang era. This example of a causal continuation from one
cycle to the next is due to Ellis \cite{elli,elal}. The singularity in 
the metric induced by the signature change is very mild. We show how in this 
case geodesics can be continued through the crunch in a completely 
smooth manner.

\section{ Cosmological Perturbation Theory in a Closed Universe}
In this section we first study the equations which govern the time 
evolution of radiation and matter density perturbations in a closed 
universe. We then  determine the power spectrum, 
which we use to decide at which length scale the density 
perturbations might first lead to the formation of objects (e.g. 
galaxies or black holes). We finally use these results to argue how
the entropy of the present cycle may have been generated.

\subsection{Time Evolution of Density Perturbations}
To describe the time evolution of density perturbations we use gauge
invariant linear cosmological perturbation theory 
(see e.g. \cite{durr}). Assuming adiabatic perturbations and neglecting 
anisotropic stresses, the evolution of the gauge
invariant density perturbation variable $D$ is governed by the equation
	\be\label{pert}
	\ddot{D}-(\nabla^2 + 3)c_{s}^{2} D+(1+3c_{s}^{2}-6\omega)
	\left(\frac{\dot{a}}{a}\right)\dot{D}
	\ee
	\[-3 \left\{\omega\left(\frac{\ddot{a}}{a}\right)
	-3\left(\frac{\dot{a}}{a}\right)^{2}(c_{s}^{2}-\omega)+(1+\omega)
	\frac{4}{3}\pi G \rho a^{2}\right\}D=0.\]
In a universe which consists of matter and radiation, 
$\omega=p/\rho=(1/3)(1+a/a_{eq})^{-1}$, 
$c_{s}^{2}=\dot{p}/\dot{\rho}
=(1/3)(1+3a/4a_{eq})^{-1}$, where $c_{s}$ is the sound velocity. A dot 
indicates
derivatives with respect to conformal time $t$ and $a_{eq}$ is the scale 
factor when $\rho_{rad}=\rho_{mat}$. 
Two cases of particular interest are dust ($\omega=
c_{s}^{2}=0, ~ a_{eq}=0$) and  radiation 
($\omega=c_{s}^{2}=1/3, ~ a_{eq}=\infty$).

Expanding $D$ in terms of scalar harmonic functions on ${\bf S}^3$, as 
described in 
Appendix~\ref{A1}, leads to the following equation for the gauge invariant
density perturbation amplitude for the wavenumber $l\in \{0,1,2,...\}$:
	\[\ddot{D}_{l}(t)+(l(l+2) - 3)c_{s}^{2} D_{l}(t)+(1+3c_{s}^{2}-6\omega)
	\left(\frac{\dot{a}}{a}\right)\dot{D}_{l}(t)\]
	\be\label{pert1}
	\quad -3 \left\{\omega\left(\frac{\ddot{a}}{a}\right)
	-3\left(\frac{\dot{a}}{a}\right)^{2}(c_{s}^{2}-\omega)+(1+\omega)
	\frac{4}{3}\pi G \rho a^{2}\right\}D_{l}(t)=0.
	\ee
For most of the sequel we omit the index $l$ which distinguishes the different 
eigenfunctions of $\nabla^2$ on ${\bf S}^{3}$.

In the following subsections we solve equation 
(\ref{pert1}) in some cases of special interest. We  then use our results 
 to  derive the power spectrum.

\subsubsection{Radiation Density Fluctuations in a Radiation 
Universe}
\begin{figure}
        \begin{center}
        \mbox{
                \psfig{file=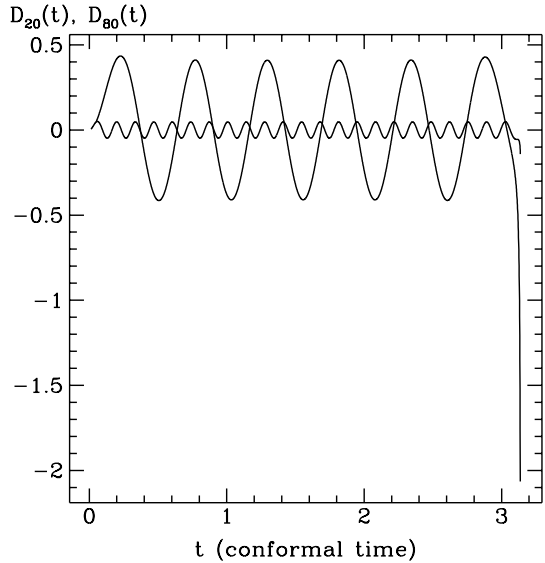,width=7.5cm}
             }
        \end{center}
\caption{\label{fig1}
The time evolution of $D_{l}(t)$ for radiation density perturbations in
a radiation dominated universe (in arbitrary units). The amplitude $D_{l}(t)$
is shown for the scales $l=20$ and $l=80$. Very close to the big crunch,
the divergence due to $\cot t$ takes over.}
\end{figure}

At early stages of expansion and at the end of the collapsing phase, the 
universe will consist of pure  
radiation, i.e.,  all matter will be relativistic. Therefore, this case is 
important for each hypothetical previous cycle, whether it 
entered the matter dominated era or it was always radiation dominated.
For radiation, where $\omega=c_{s}^{2}=1/3$, equation (\ref{pert1}) reduces to
	\be\label{pert2}
	\ddot{D}+\left\{\frac{(l(l+2)-3)}{3}-\left(\frac{\ddot{a}}{a}\right)
	-\frac{16}{3}\pi G \rho a^{2}\right\}D=0.
	\ee
Inserting the solution of the Friedmann equation for the scale factor of 
a radiation dominated 
universe, which is $a(t)=a_{\max}\sin t$ with 
$a_{\max}=(8\pi G \rho a^{4}/3)^{1/2}$, we obtain
	\be\label{rrdu}
	(\sin^{2}t)\ddot{D}+\left( \frac{l(l+2)}{3}\sin^{2} t - 2\right) D = 0,
	\qquad t\in [0,\pi].
	\ee
The solution of this equation is given by
	\be\label{rrdus}
	D(t)=\sin^{2}t\left(\frac{1}{\sin t}\frac{d}{dt}\right)^{2}\left[
	c_{1} \exp\{i\sqrt{a_{l}}t\}+ c_{2} \exp\{-i\sqrt{a_{l}}t\}\right]
	\ee
with $a_{l}=l(l+2)/3$ and $l\not= 0$ (see \cite{kamk}). 
We are only interested in real solutions for positive integers $l$, in which 
case $D(t)$ can be written in the form
	\bea\label{rrdus1}
	D(t)=& & c_{1}\left[ \sqrt{a_{l}}\cot t \sin(\sqrt{a_{l}}t)-a_{l}
	\cos(\sqrt{a_{l}}t)\right]  \nonumber \\
	&+& c_{2}\left[ \sqrt{a_{l}}\cot t \cos(\sqrt{a_{l}}t)-a_{l}
	\sin(\sqrt{a_{l}}t)\right].
	\eea
This solution is plotted in Fig.~\ref{fig1} for $l=20$ and $l=80$.
Obviously the amplitude $\cot t$ diverges at the big bang
and at the big crunch where $t=0$ and $t=\pi$ respectively. Since we 
assume that the fluctuations are created at some time $t_i >0$ 
after the big bang, the divergence at
$t=0$ is not a problem. Apart from its oscillation with frequency 
$\sqrt{a_l} \sim l$, the amplitude of 
 density fluctuations is  approximately constant for most of
the cycle, but diverges close to the crunch like
$D_{l}(t)\propto l(\pi-t)^{-1}$.

\subsubsection{Matter Density Fluctuations in a Radiation Dominated Universe}
For dust ($\omega=c_{s}^{2}=0$) equation (\ref{pert1}) reduces to
	\be\label{dust}
	\ddot{D}+\frac{\dot{a}}{a}\dot{D}-4\pi G\rho_{mat} a^{2} D=0.
	\ee
As long as the universe expands, $\dot{a}$ is positive and hence the second
term in this equation acts as a damping term. This term vanishes
at maximum expansion and turns into a stimulation 
when the universe contracts. Therefore we expect the
growth of  fluctuations to become substantially enhanced during the
contraction phase.

Inserting the scale factor $a(t)=a_{max}\sin t$ for the radiation dominated 
universe, we obtain
	\be\label{mrdu}
	\sin t \ddot{D}+\cos t \dot{D} - \mu D = 0,\qquad t\in [0,\pi],
	\ee
where $\mu=4\pi G \rho_{mat}a^3=(3/2)(a_{max}/a_{eq})$ and 
$a_{max}=(8\pi G \rho a^{4}/3)^{1/2}$. With the 
substitution $x=\sin t$ this equation leads to 
	\be\label{heun1}
	x(x-1)(x+1)D''+(2x^{2}-1)D'+\mu D=0.
	\ee

This equation is a special case of Heun's differential equation. 
For $|x|<1$ one solution
	\[D_{a}(x)=1+\sum_{n=1}^{\infty}c_{n}x^{n}\]
of (\ref{heun1}) is a convergent power series with 
	\bean
	c_{1}&=&\mu \\
	c_{n+1}&=&\frac{n(n-1)}{(n+1)^{2}}c_{n-1}+\frac{\mu}{(n+1)^{2}}c_{n}.
        \eean

\begin{figure}
        \begin{center}
        \mbox{
                \psfig{file=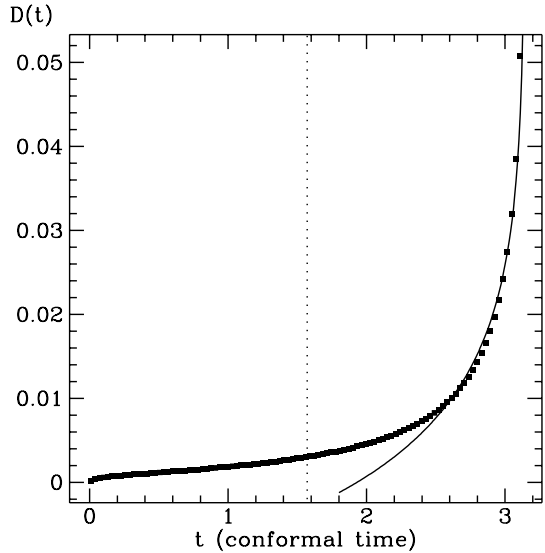,width=7.5cm}
             }
        \end{center}
\caption{\label{fig2}
The time evolution of $D(t)$ for matter density perturbations in a
radiation dominated universe (in arbitrary units). For comparison, a
logarithmically divergent fit is shown close to the collaps (solid line).
The vertical dotted line indicates $t_{max}$, the time when the cycle
reaches its maximum expansion.}
\end{figure}

A numerical solution of (\ref{mrdu}) is shown in Fig.~\ref{fig2}.

Let us discuss the behavior of  matter density  
fluctuations more closely during the different epochs of a cycle.
At early times, when $t\ll 1$, equation (\ref{mrdu}) simplifies to
	\be\label{early}
	t \ddot{D} + \dot{D} -\mu D=0
	\ee
which has a solution in terms of Bessel functions:
	\[D(t)=c_{1}J_{0}(2i\sqrt{\mu t})+c_{2}Y_{0}(2i\sqrt{\mu t})\approx
	\tilde{c_{1}}+\tilde{c_{2}}ln(2\sqrt{\mu t}).\]
Neglecting logarithmic growth, these fluctuations are 
approximately constant for $t\ll 1$. Fig.~\ref{fig2} shows, 
that the logarithmic growth for small $t$ is  a good 
approximation up to $t\approx 1/10$ which means that all scales with 
$l\geq 10$ or so enter the horizon during  this era,
 where we can consider the fluctuations to be approximately constant.
This is the well known M\'ez\`aros effect:
matter fluctuations do not grow in a flat radiation dominated universe.
Close to maximum expansion ($t=\pi/2 \pm \ep,\ |\ep| \ll 1$),
equation (\ref{mrdu}) reduces to
	\[{d^2D\over \ep^2} -\ep{dD\over\ep} -\mu D =0\]
The solution can be written
in terms of the confluent hypergeometric function:
\be    D(\ep)=\ep^{-1/2}e^{\ep^{2}/4}Y\left({\mu\over 2} -
	\frac{1}{4},\frac{1}{4},\frac{-\ep^{2}}{2}\right) ~, \label{Dep} \ee
with
	\[Y(k,m,x)=c_{1}M_{k,m}(x)+c_{2}M_{k,-m}(x)\]
	\[M_{k,m}(x)=x^{1/2+m}e^{-x/2} F(1/2+m-k,2m+1,x)\]
	\[F(a,b,x)=1+\sum_{n=1}^{\infty} \frac{a(a+1)\ldots (a+n-1)x^{n}}
	{b(b+1)\ldots (b+n-1)n!}.\]
Near maximum expansion, small $|\ep|$, $D$ grows nearly exponentially, like
in a non--expanding universe.

Close to the big crunch, where $t=\pi - \ep$, we again obtain equation
(\ref{early}), replacing $t$ and dots   by $\ep$ and derivatives
with respect to $\ep$. This reflects the symmetry of the closed universe 
between (big bang, $t$) and (big crunch, $-t$) as long as the entropy remains
unchanged. Therefore, close to the big crunch $D$ diverges logarithmically:
	\[D(t)\propto \ln \left(\frac{1}{2\sqrt{\mu (\pi -t)}}\right).\]
This solution is valid until the particles become relativistic
around the de--con\-fine\-ment phase transition where $T\cong 100\:MeV$,
from where on we  have to consider pure radiation density fluctuations.

\subsubsection{Matter Density Fluctuations in a Matter Dominated Universe}
In the case of a matter dominated universe, the scale factor is 
given by $a(t)=(a_{\max}/2)(1-\cos t)$, 
where $a_{\max}=8\pi G \rho a^{3}/3$ and hence (\ref{dust}) reads
	\be\label{dusta}
	(1-\cos t)\ddot{D}+(\sin t) \dot{D}-3D=0,\qquad t\in [0,2\pi ],
	\ee
with the well known solution (see \cite{wein})
	\be\label{mmdus}
	D(t)=c_{1}\left[\frac{5+\cos t}{1-\cos t}
	-\frac{3t\sin t}{(1-\cos t)^{2}}\right]+c_{2}
	\left[\frac{\sin t}{(1-\cos t)^{2}}\right]
	\ee
At early times, equation (\ref{dusta}) is approximately given by
$t^{2}\ddot{D}+2t\dot{D}-6 D=0$, such that  $D\propto t^{2}$ 
or $D\propto t^{-3}$, for the growing and decaying mode respectively.
Close to the collapse, for $t=\pi -\ep$, equation (\ref{dusta}) 
reduces again to
	\[\ep^{2}\frac{d^{2}}{d\ep^{2}}D+2\ep\frac{d}{d\ep}D-6 D=0,\]
but the growing and decaying modes are interchanged. Now the growing mode
solution is given by $D=D_0\ep^{-3}=D_{0} (2\pi -t)^{-3}$. At 
maximum expansion, the damping term  vanishes and the evolution of the
fluctuations around $t=\pi$ is described by exponential growth or decay.

\subsubsection{Composite Model}
To construct a more realistic model where the scale factor is not 
only determined by a single matter or radiation background, we
now assume a simple composite model, where the energy density of the 
universe is given by
	\[\rho(a)=\frac{\rho_{eq}}{2}\left[\left(\frac{a_{eq}}{a}\right)^{3}
	+\left(\frac{a_{eq}}{a}\right)^{4}\right].\]
The first expression on the right hand side represents the 
$a^{-3}$ behavior of the matter 
density and the second term reflects the $a^{-4}$ behavior of radiation 
density. The solution of the Friedmann equation (\ref{fried}) in this case is
	\be\label{scalef}
	a(t)=\sqrt{\Delta}\sin \left(t-\arcsin \left(\frac{\tilde{a}}
	{\sqrt{\Delta}}\right)\right)+\tilde{a}= a_{eq}(\al
	\sin{t}+\al^{2}\sin^{2}\frac{t}{2}),
	\ee
where $\al=a_{eq}/a_0$ with  $a_{0}=(4\pi G \rho_{eq}/3)^{-1/2}$  and
$\tilde{a}={1\over 2}\al^2a_{eq}$,
$\Delta=\al^2a_{eq}^2+\tilde{a}^{2}$.

Furthermore, we find from (\ref{scalef}) that 
	\[t_{eq}=\arcsin(\Delta^{-1/2}(a_{eq}-\tilde{a}))
	+\arcsin(\tilde{a}\Delta^{-1/2}),\] 
	\[t_{max}=\pi/2+\arcsin(\tilde{a}\Delta^{-1/2}),\] 
	\[a_{max}\equiv a(t_{max})=(\Delta^{1/2}+\tilde{a})=\frac{1}{2}
	a_{eq}\alpha\left(\alpha+\sqrt{4+\alpha^{2}}\right).\]
We can use $\alpha$ as a parameter which determines the duration of the 
matter dominated epoch in a closed universe containing matter and radiation.
For $\al\ll 1$, 
$a_{\max} \approx \al a_{eq} \ll a_{eq}$ and the universe never becomes matter
dominated. For $\al\gg 1$, $a_{\max} \approx\al^2a_{eq}\gg a_{eq}$ and
$t_{eq}\ll t_{\max} \approx \pi$; the universe experiences a long matter 
dominated era.

\begin{figure} 
        \begin{center}
        \mbox{
                \psfig{file=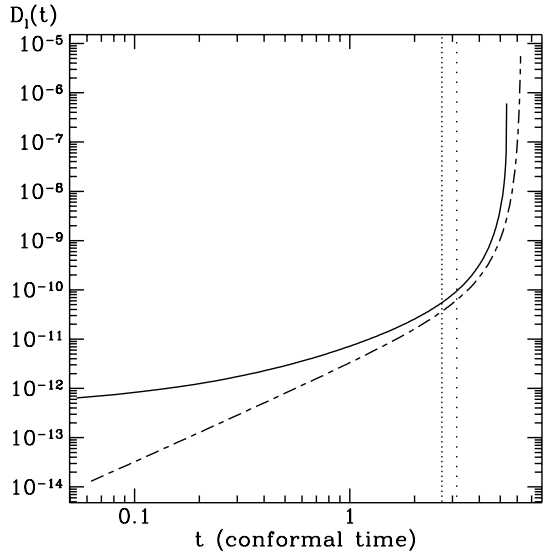,width=7.5cm}
             }
        \end{center}
\caption{\label{fig3}
The amplitude $D(t)$ of matter density perturbations in the composite
model (in arbitrary units). The  solid line shows $D(t)$ for a cycle with
a short matter dominated phase ($\alpha=4$). The dashed line shows
$D(t)$ for a cycle with a long matter dominated epoch ($\alpha = 1000$,
$a_{max}\gg a_{eq}$). The left and right vertical dotted lines indicate the
time of maximum expansion of the cycle for $\alpha=4$ and $\alpha=1000$
respectively.}
\end{figure}

For radiation density perturbations in this
composite model, equation (\ref{pert1}) yields
	\be\label{comp1}
	a^{2}(t)\ddot{D}(t)+\left\{a^{2}(t)\left(\frac{l(l+2)-3}{3}\right)
	-a(t)\ddot{a}(t)-4a_{eq} \tilde{a}\right\}D(t)=0,
	\ee
and for matter density perturbations we obtain
	\be\label{comp2}
	a(t)\ddot{D}(t)+\dot{a}(t)\dot{D}(t)-3\tilde{a}D(t)=0,
	\ee
with $a(t)$ given by (\ref{scalef}). We are particularly interested in
equation (\ref{comp2}). Numerical solutions for the growing mode of a short
and long matter dominated phase are shown in Fig.~\ref{fig3}.

\subsection{The Power Spectrum}
\subsubsection{The Harrison Zel'dovich initial spectrum}
The power spectrum $P(l,t)$ determines the scaling
behavior of perturbations at a given time $t$. It is defined by
	\be
	P(l,t)\equiv |D_{l}(t)|^{2},
	\ee
where $D_{l}(t)$ is a solution of equation (\ref{pert1}).  
To determine the power spectrum, we have to specify the $l$--dependence of 
the initial amplitudes, $D_l(t_{in})$. A preferred such choice,
which we also adopt here, is the scale invariant 
or Harrison--Zel'dovich spectrum \cite{haze}. The power spectrum is called
Harrison--Zel'dovich if the 
variance of the mass fluctuation
on horizon scales $R_{H}=\int_{0}^{t}dt= t$ is constant, time independent:
	\[\langle (\de M/M)^{2}_{R_{H}}\rangle=\mbox{const.}\]
Here $\langle\cdot\rangle$ denotes the statistical average over many 
'realisations' of perturbed universes with identical statistical properties.
Since we know only one such universe, we assume that this statistical average
can be replaced by a spatial average, a kind of 'ergodic hypothesis'. 
We want to express 
$(\de M)_{R_{H}}(t)$ as a function of $D_{l}(t)$. Let us therefore identify
the (gauge invariant) density variable $D(x)$ with $(\de\rho/\rho)(x)$ and let
$l_{H}$ be the value of $l$ corresponding to the horizon size 
$R_{H}=\pi/l_{H}$. Let us denote the spherical harmonics on ${\bf S}^3$ 
by ${\cal Y}_{\bf k}$, where $\bf k$ stands for the multi--index 
$(l,j,m)$ specifying the the spherical
harmonics on ${\bf S}^3$ (see Appendix \ref{A1}). We then obtain for the
mass fluctuation inside a volume of size $R_H^3$
	\bean
	(\de M)_{R_{H}}(t)&=&\int_{V_{H}} d^{3}x\:h^{1/2}\de\rho({\bf x},t)
	=\rho\int_{V_{H}} d^{3}x\:h^{1/2}(\de\rho / \rho)  \\
	 &=&\rho\int_{V_{H}}d^{3}x\:h^{1/2}
	\sum_{{\bf k}}{\cal Y}_{{\bf k}} \left(\frac{\de\rho}{\rho}\right)
	_{\bf k}(t)
	\approx \rho V_{H} \sum_{l\leq l_{H}}\sum_{j,m}{\cal Y}_{{\bf k}} 
	({\bf x}) D_{l}(t).
	\eean
For the final approximation, we have assumed that perturbations on scales 
smaller than $\pi/l_{H}$ average to zero due to the integration over 
$V_{H}$, and that perturbations on scales larger than $\pi/l_{H}$ are 
approximately constant in a volume of size
$R_H\sim \pi/l_H$, such that integration
over $V_{H}$ just gives rise to the factor $V_{H}$ ( = volume of a three
dimensional patch of diameter $2 R_{H}$ on ${\bf S}^{3}$). 
With $M=\rho V_{H}$, we then obtain 
	\[(\de M/M)_{R_{H}}^{2}\approx\sum_{l,l'\leq l_{H}}\quad
	\sum_{(j,m),(j',m')}
	{\cal Y}_{{\bf k}}({\bf x}){\cal Y}_{{\bf k}'}^{*}({\bf x})
	D_{l}(t)D_{l'}^{*}(t)\]
and
	\bea\label{pspec}
	\langle(\de M/M)^{2}_{R_{H}}\rangle &\approx& \sum_{{\bf k},{\bf k'}
	\leq l_{H}} D_{l}(t)D^{*}_{l'}(t) \int_{{\bf S}^{3}} d^{3}x \:h^{1/2}
	{\cal Y}_{{\bf k}}({\bf x}){\cal Y}_{{\bf k}'}^{*}({\bf x})\nonumber\\
	&=& \sum_{l\leq l_{H}}\sum_{j=0}^{l-1}\sum_{m=-j}^{j}P_{l}(t)
	=\sum_{l\leq l_{H}}l^{2}P_{l}(t).
	\eea
Let $t_{l}\approx \pi/l$ denote the time when the scale $l$
crosses the horizon, $l_{H}(t_{l})=l$ . Since
$\sum_{l\leq l_{H}}l^{2}\approx l_{H}^{3}$, we see from (\ref{pspec}) that we
have to demand that 
\[ P(l,t_{l}= \pi/l) \propto l^{-3}\]
for $\langle(\de M/M)^{2}_{R_{H}}\rangle$ to be approximately constant, 
i.e.,  for a scale invariant spectrum.

The notion of a scale invariant power spectrum  can now 
be used to compare the time evolution of density perturbations on 
different length scales. We want to investigate, which length scale 
 collapses first. It is the scale
at which the variance  of the mass perturbation first grows
 of order unity. At that time, linear perturbation theory breaks down 
and  we expect matter perturbations to form gravitationally bound objects.

\subsubsection{The Final Power Spectra}
Up to an overall constant,  the scale invariant spectrum is 
determined  by the $t$- and $l$-dependence of the density 
perturbations $D_{l}(t)$ and  by the requirement that 
$D_{l}(t_{l}=\pi/l)$ is proportional to $l^{-3/2}$.
Then of course $P(l,t_{l})$ is proportional to $l^{-3}$ and the 
variance of the mass perturbation is approximately constant. 

\begin{figure} 
        \begin{flushleft} 
        \mbox{   
                (A)\psfig{file=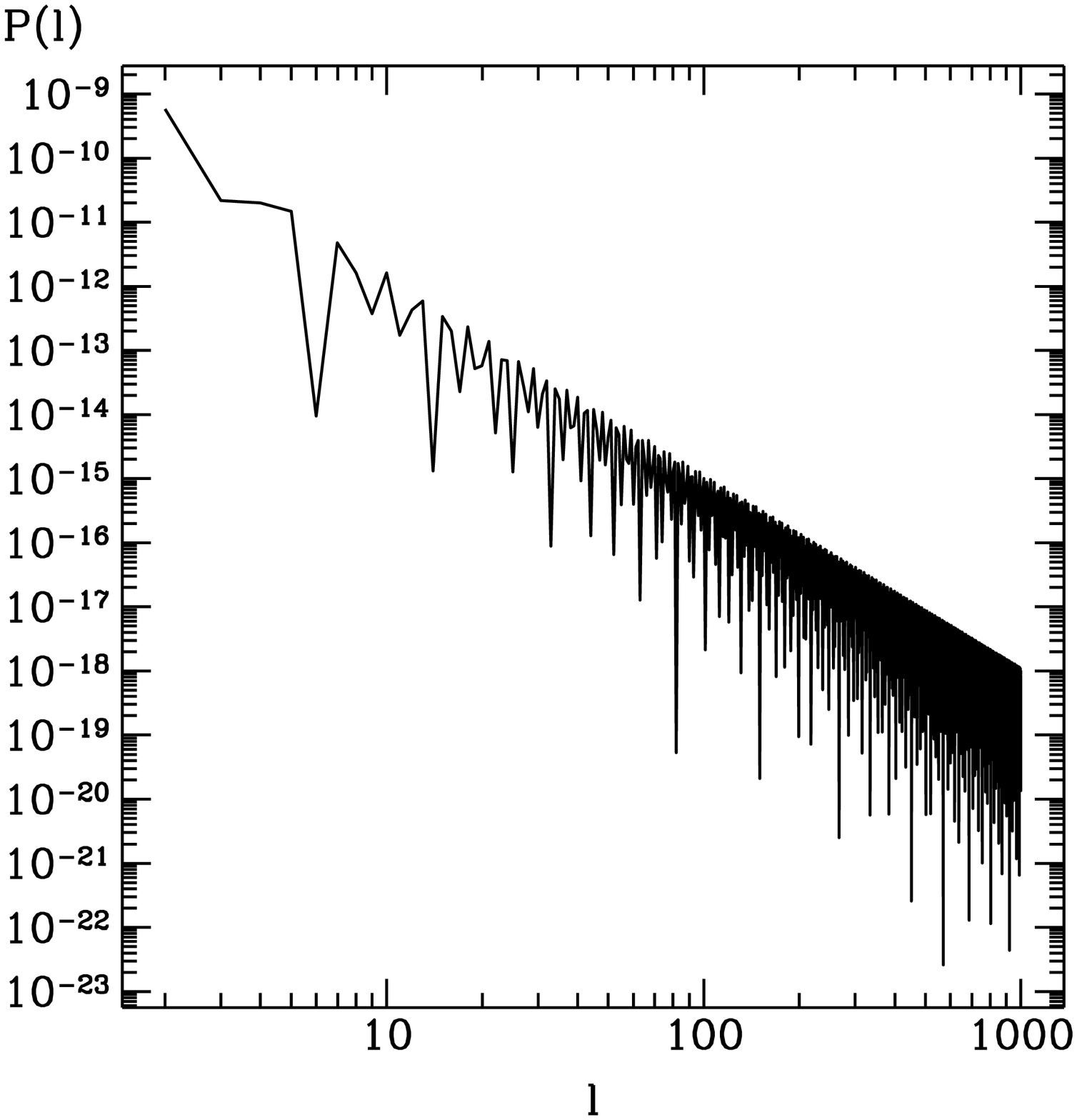,width=7.5cm}
                (B)\psfig{file=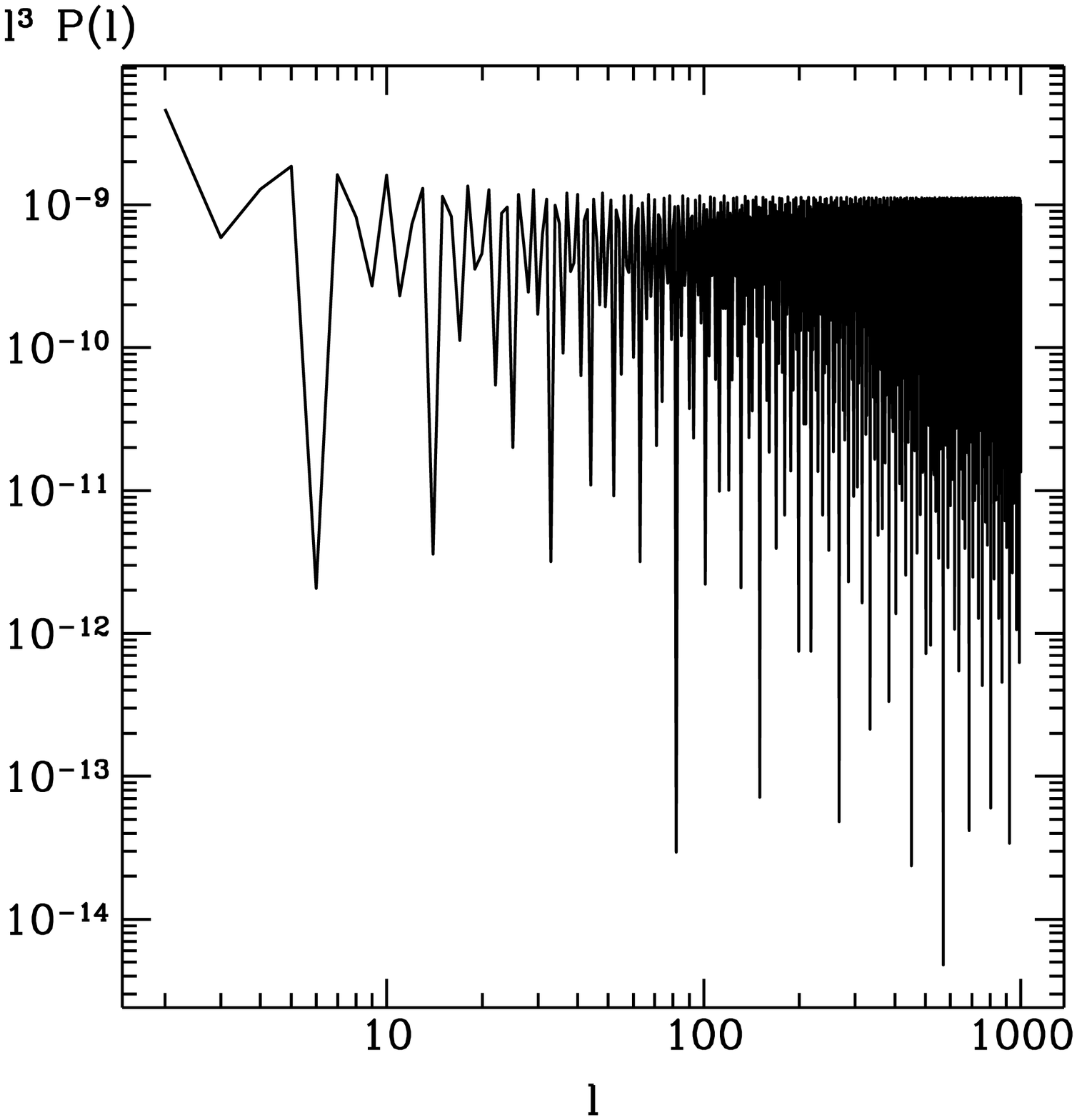,width=7.5cm}
             } 
        \end{flushleft} 
\caption{\label{fig4} 
(${\bf A}$): The power spectrum $P(l)$ as a function of $l$ for
radiation density perturbations in the
radiation dominated universe (in arbitrary units). $P(l)$ is given at
a time, when all scales $l\geq 2$
are inside the horizon. (${\bf B}$): The corresponding mass fluctuation
$l^{3} P(l)$ as a function of $l$.}
\end{figure}

Let us first apply this procedure to the case of radiation
density fluctuations in the radiation dominated epoch. From solution
(\ref{rrdus1}) we find that for a fluctuation which crosses the horizon 
at times
$t\ll \pi$, i.e. at times much smaller than the time of the big crunch,
the maximum amplitude of $D_{l}$ is approximately constant (since the term 
containing $\cot t$ is small) and therefore $c_{1}$ and $c_{2}$ must be 
proportional to $l^{-7/2}$ (since $a_{l}$ is proportional to $l^{2}$) to obtain
the required $l^{-3/2}$ behavior of $D_{l}(t_{l})$. 
Close to the crunch, when $t\to \pi$,  the expression containing $\cot t$
diverges as $(\pi-t)^{-1}$. But all scales $l\geq 2$ enter the horizon at 
times $t_{l}\leq \pi/2$, and therefore this divergence is only relevant 
for the mode $l=1$, which enters the horizon at the big crunch.
Of course the scales which enter the horizon 
already for $t\ll \pi$, also begin to grow as $(\pi -t)^{-1}$, when $t$
approaches $\pi$. Disregarding the $l=1$ mode we thus obtain
	\be
	D_{l}(t)\propto \left\{\begin{array}{ll} 
	l^{-3/2} , & t\ll \pi\\
	l^{-3/2}(\pi-t)^{-1}, & t\to\pi ,
	\end{array}\right. 
	\ee
or equivalently
	\be
	P(l,t)\propto \left\{\begin{array}{ll} 
	l^{-3} , & t\ll \pi\\
	l^{-3}(\pi-t)^{-2}, &t\to \pi
	\end{array}\right. 
	\ee
for radiation perturbations in the radiation dominated era.  
At late times, close to the crunch, we can therefore  approximate
the power spectrum by $P(l,t)\cong c^{2}l^{-3}(\pi-t)^{-2}$.
This power spectrum obviously takes its maximum for the smallest
value of $l$, and the induced mass fluctuations $l^{3}P(l)$ are 
independent  of  scale (see Figs.~\ref{fig4}(A) and \ref{fig4}(B)).

As the next example we consider matter density perturbations in the radiation
dominated era. We have found that for $t< 1/10$ they show logarithmic growth
which we approximate by a constant.
To obtain a scale invariant spectrum we therefore have to require
$D_l\propto l^{-3/2}$ and hence again $P(l)\propto l^{-3}$. Only the largest scales
which enter the horizon close to or after maximum expansion do not satisfy this
proportionality since we can not assume the corresponding density 
fluctuations to
be approximately constant. Since the density perturbations grow with a certain
power of $t$, the slope of the spectrum will decrease towards the largest 
scales. Matter density fluctuations are a special case of the composite
model, when the cycle never reaches the matter dominated phase, $\al\ll 1$.
The numerically determined power spectrum $P(l)$ and the mass fluctuation 
$l^3P(l)$ are shown in curves A and B of Fig.~\ref{fig5}.

Now we determine the scale invariant power spectrum for matter 
density fluctuations  in the matter dominated era.
For small scales, which enter the horizon early, where $D_{l}(t)\propto 
A_l t^{2}$ scale invariance requires $A_l\propto l^{1/2}$. However when 
the cycle approaches it's
maximum expansion for $t\to \pi$, the damping term is smaller
and $D_{l}$ grows faster, say $D\propto t^{p}$ with $p>2$ (around
$t_{max}=\pi$ there is actually exponential growth, i.e. $p$ diverges for 
$l=1$) and we need $c\propto l^{-3/2+p}$. Towards the crunch,
$D$ is proportional to $(2\pi -t)^{-3}$. We finally obtain roughly
 the following $l$-dependence of the power spectrum 
	\[P(l,t)\propto\left\{ \begin{array}{lll}
	l(2\pi-t)^{-6} ,\qquad &l\gg 1 , & 0\ll 2\pi-t\ll 1 \\
	l^{2p-3}(2\pi-t)^{-6},\qquad &l\cong 1,\quad p >2,
	& 0\ll 2\pi-t\ll 1 \end{array}\right. \]
For small $l$ (large scales), the slope of $P(l)$ is bigger than 
one (in a log-log diagram), 
where as for large $l$ (small scales), the slope of $P(l)$ is equal to one.
This behavior is equivalent to the special case of the composite
model with a long matter dominated epoch, $\al \gg 1$. Therefore $P(l)$
and $l^3P(l)$ in the matter dominated universe are very similar to the
power spectra shown in Fig.~\ref{fig5} (E) and (F) for $l<1000$. This figure
actually shows a composite model with $\al= 1000$. Therefore, scales with
$l>1000$ enter the horizon still in the radiation dominated era and thus
do not represent this case. In a pure matter universe, there is no bend
in the power spectrum for $l\gg 1$.

Finally we approximate the power spectrum for the realistic 
composite model. The power spectrum in this case is composed of
three parts. At late times, when all scales are already inside the horizon, 
we obtain
($t_{eq}=\pi/l_{eq}$ denotes the time when $\rho_{mat}=\rho_{rad}$)
	\be\label{power4}
	P(l)\propto\left\{\begin{array}{ll}
	l^{-3},\qquad & l\gg l_{eq} \\
	l ,&1\ll l \ll l_{eq} \\
	l^{2p-3},&l\cong 1,\qquad p >2.
	\end{array} \right. 
	\ee
Here, the $l^{-3}$-dependence is due to fluctuations which enter the horizon
already during the radiation dominated epoch like in the flat universe. 
The maximum of the power 
spectrum is expected at $l\approx l_{eq}$. In Fig.~\ref{fig5}, (A)--(F)  some examples
for the power spectrum and the corresponding mass fluctuation are plotted. 
If the cycle has a long matter
dominated epoch, the largest scale $l=1$ enters the horizon soon after
maximum expansion of the universe. This is different if the cycle does 
not reach the matter domination. Then the largest scale 
enters the horizon very close to the 
crunch and it will be the scale $l=2$  which enters the horizon 
close to maximum expansion. 
\begin{figure}
        \begin{flushleft}
        \mbox{
                (A)\psfig{file=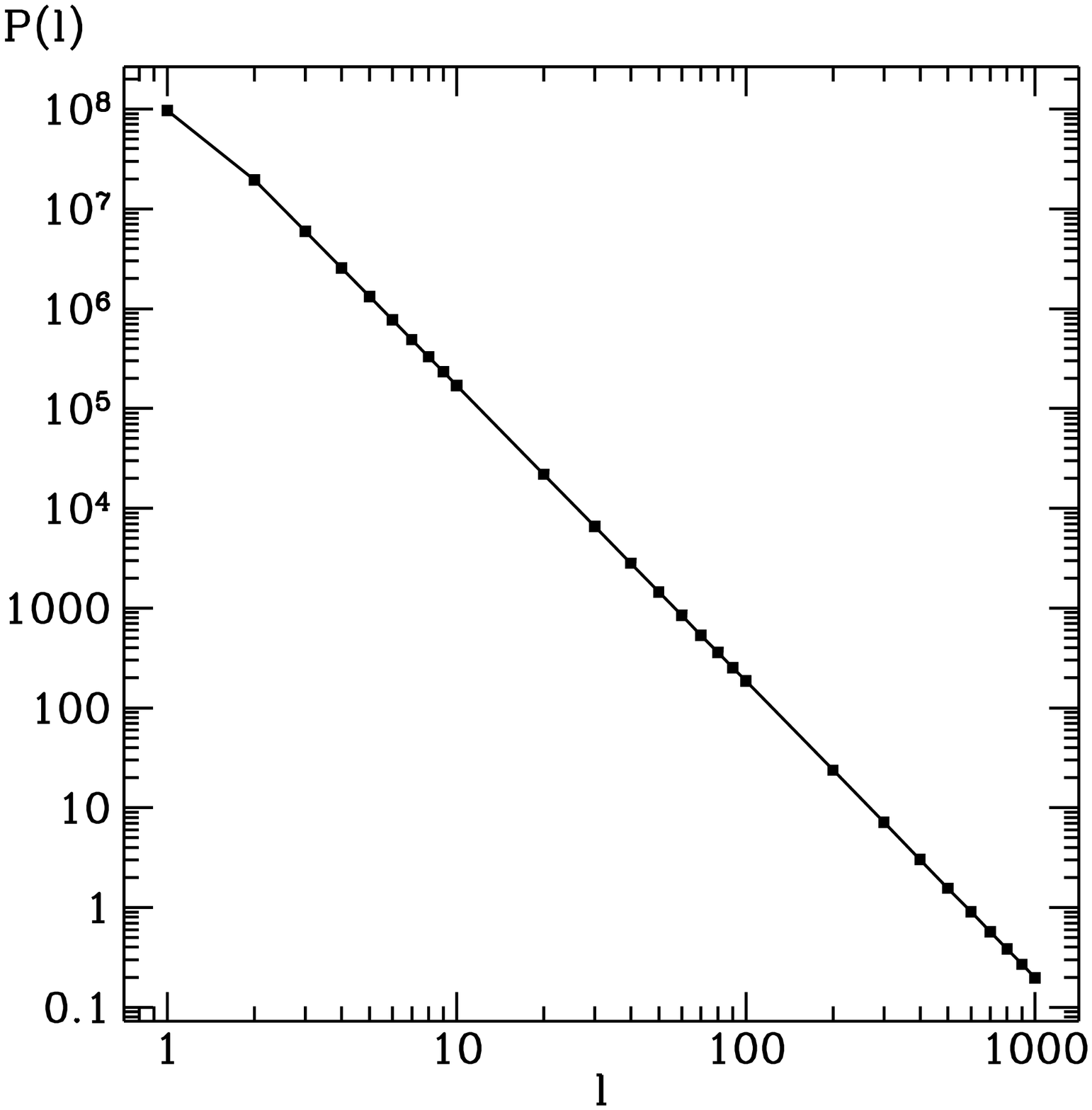,width=7.5cm}
                (B)\psfig{file=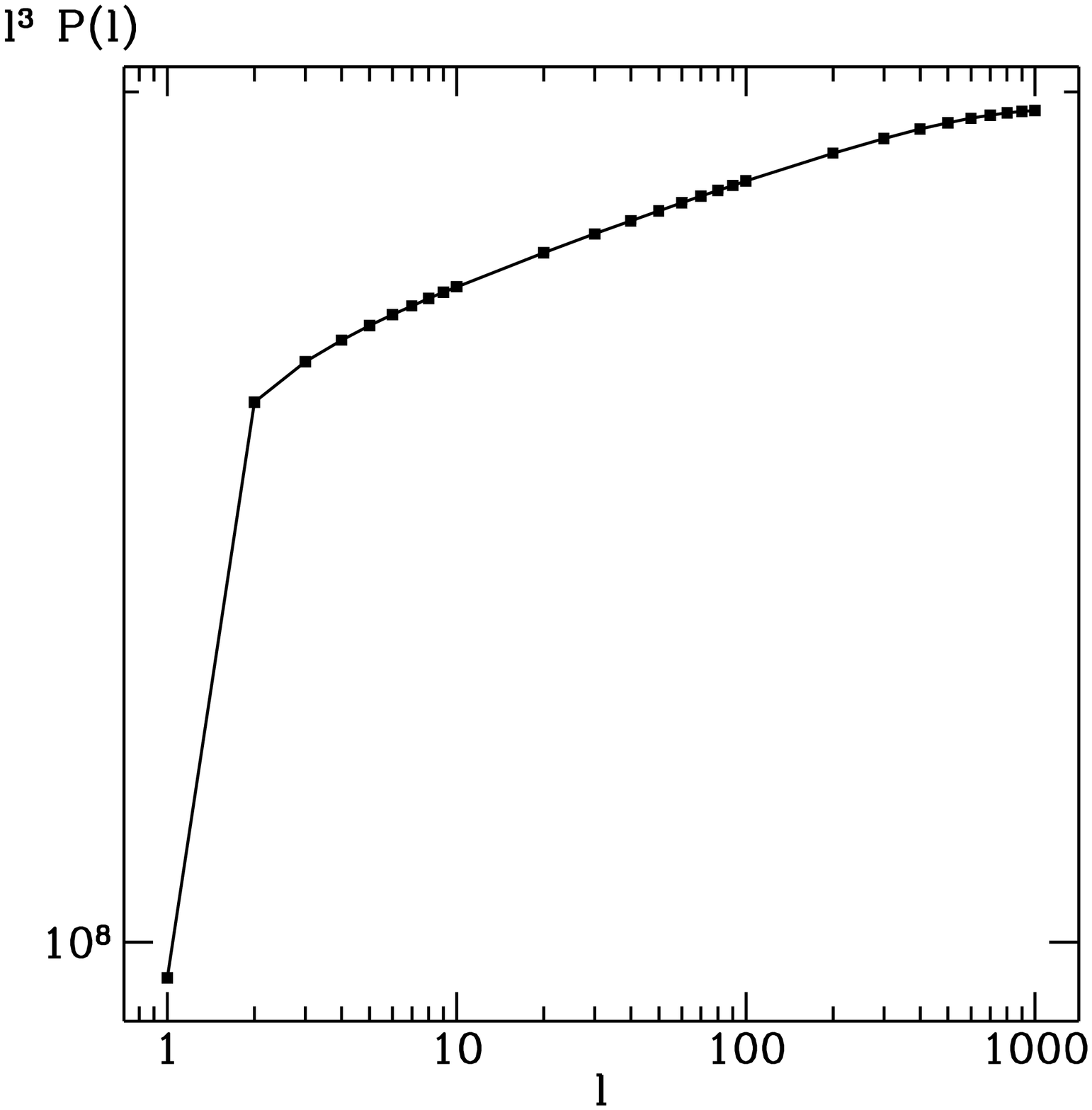,width=7.5cm}
             }
        \end{flushleft}
        \begin{flushleft}
        \mbox{
                (C)\psfig{file=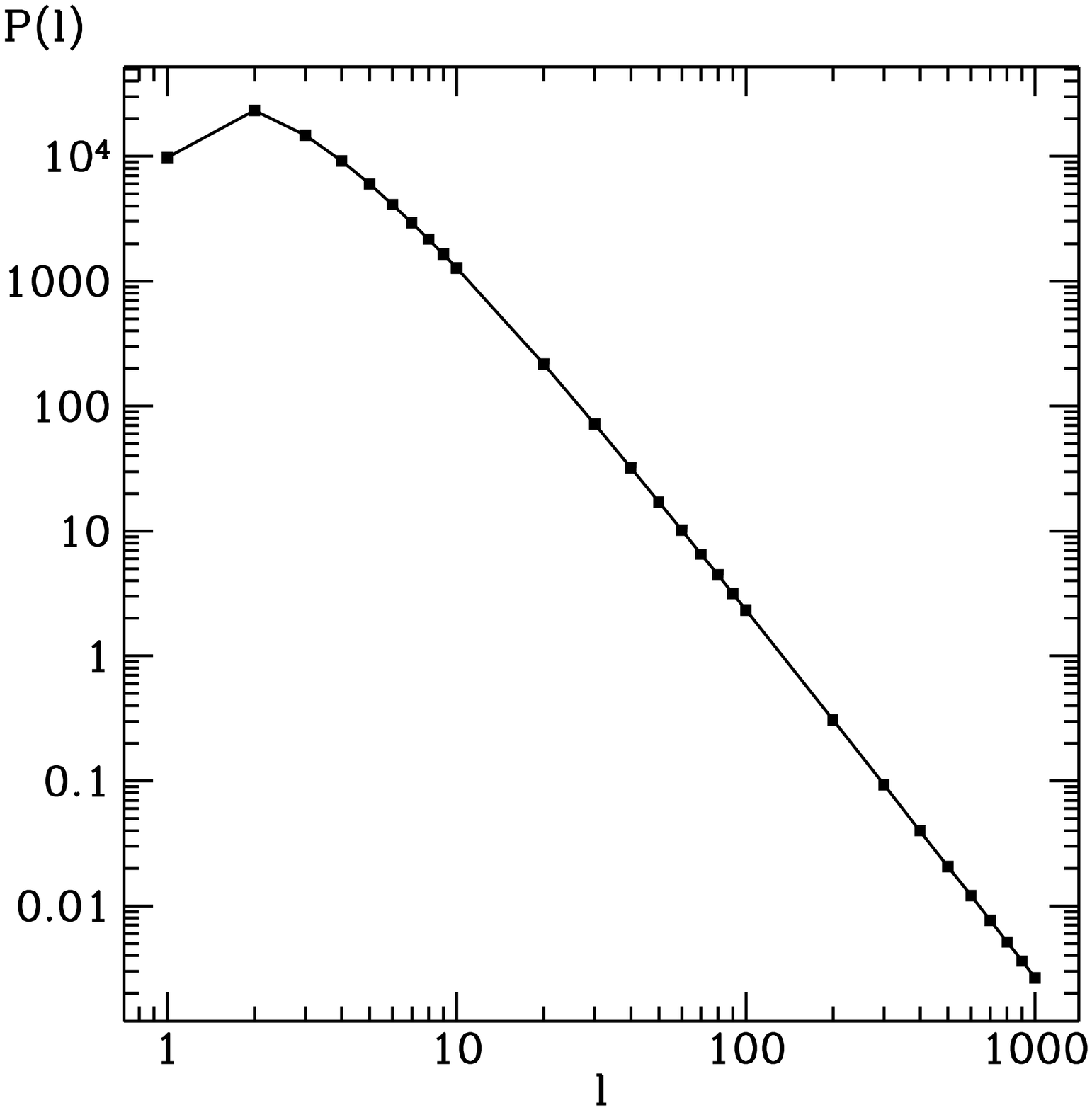,width=7.5cm}
                (D)\psfig{file=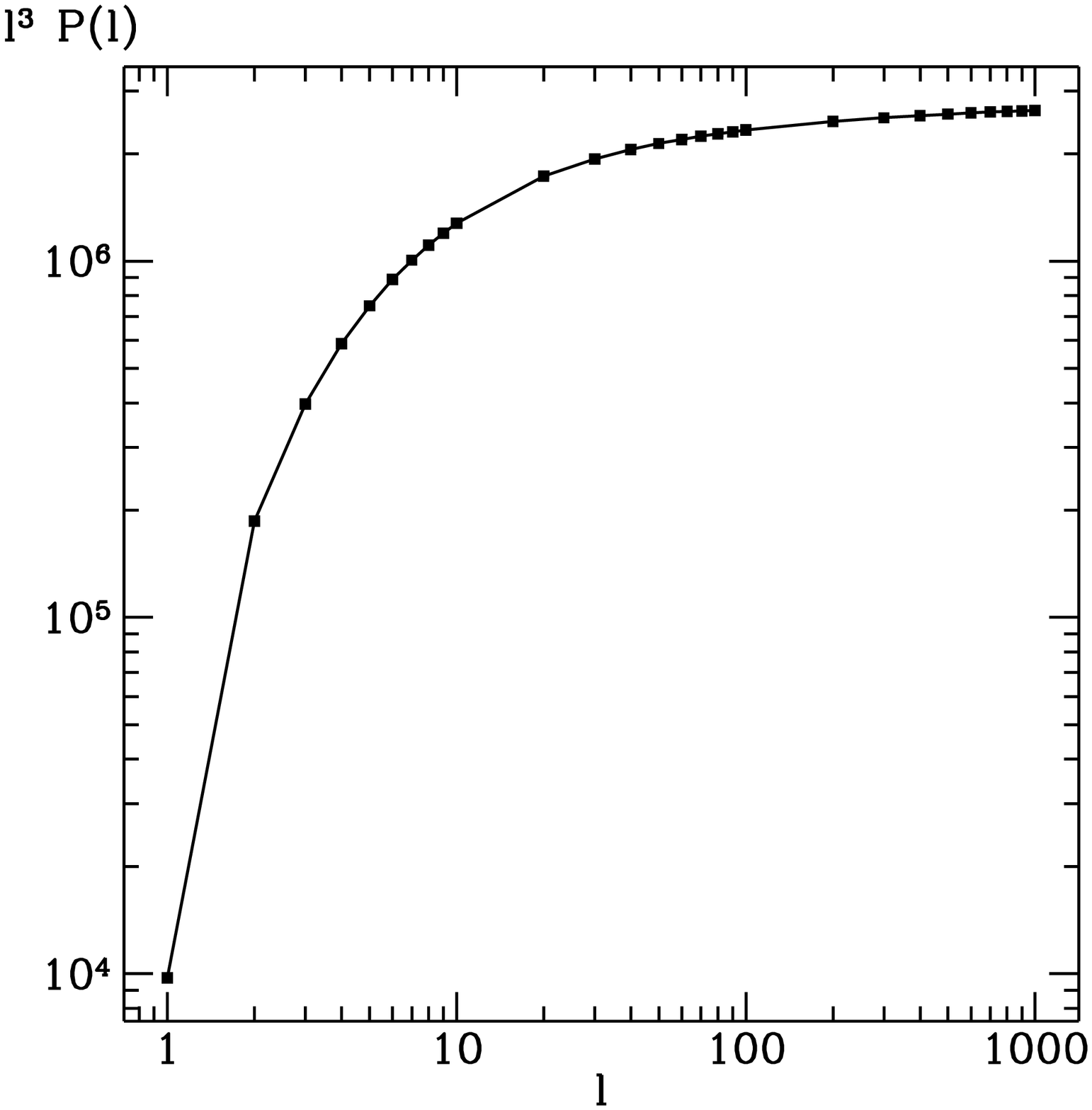,width=7.5cm}
             }
        \end{flushleft}
\end{figure}
\begin{figure}
        \begin{flushleft}
        \mbox{
                (E)\psfig{file=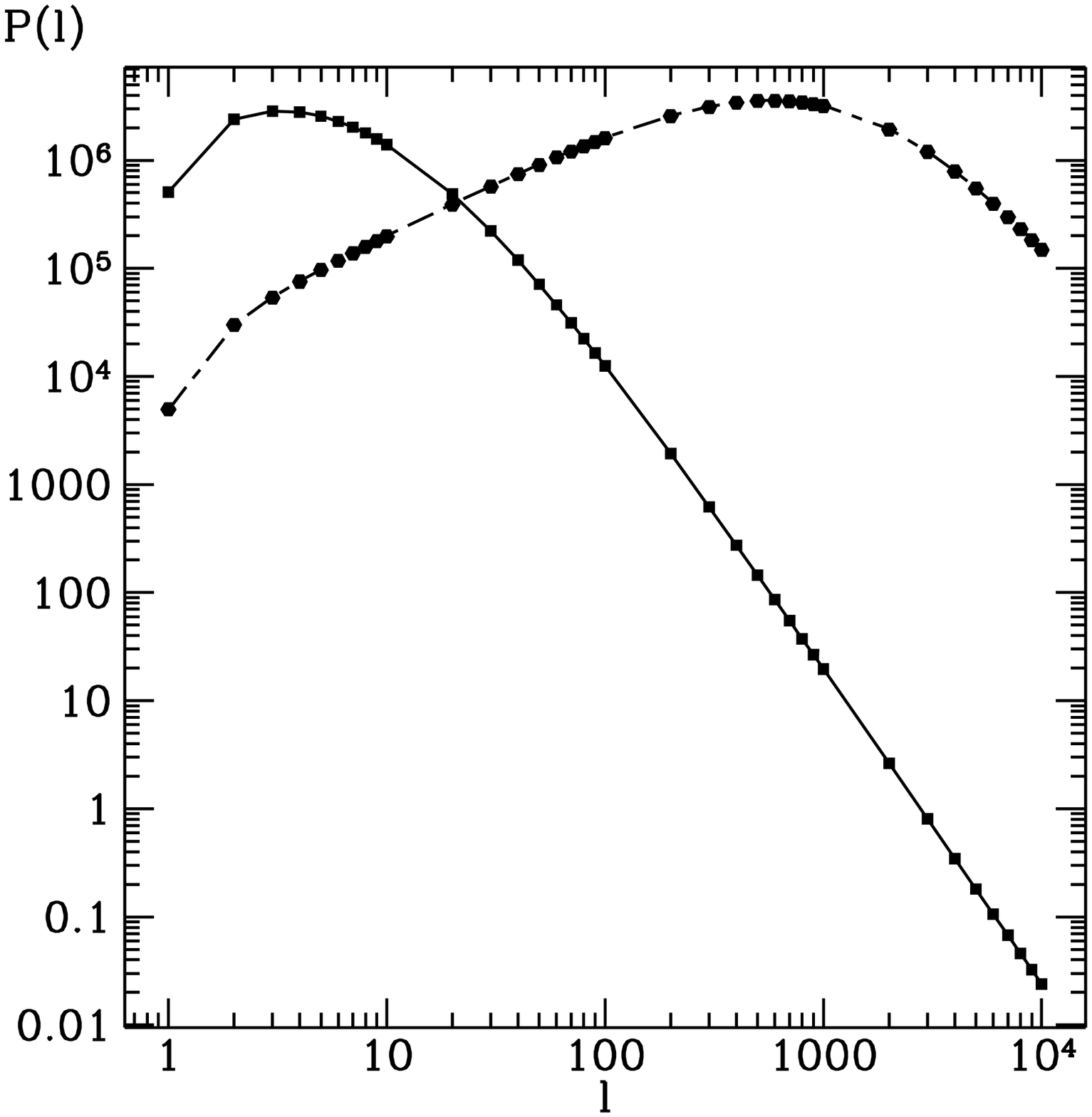,width=7.5cm}
                (F)\psfig{file=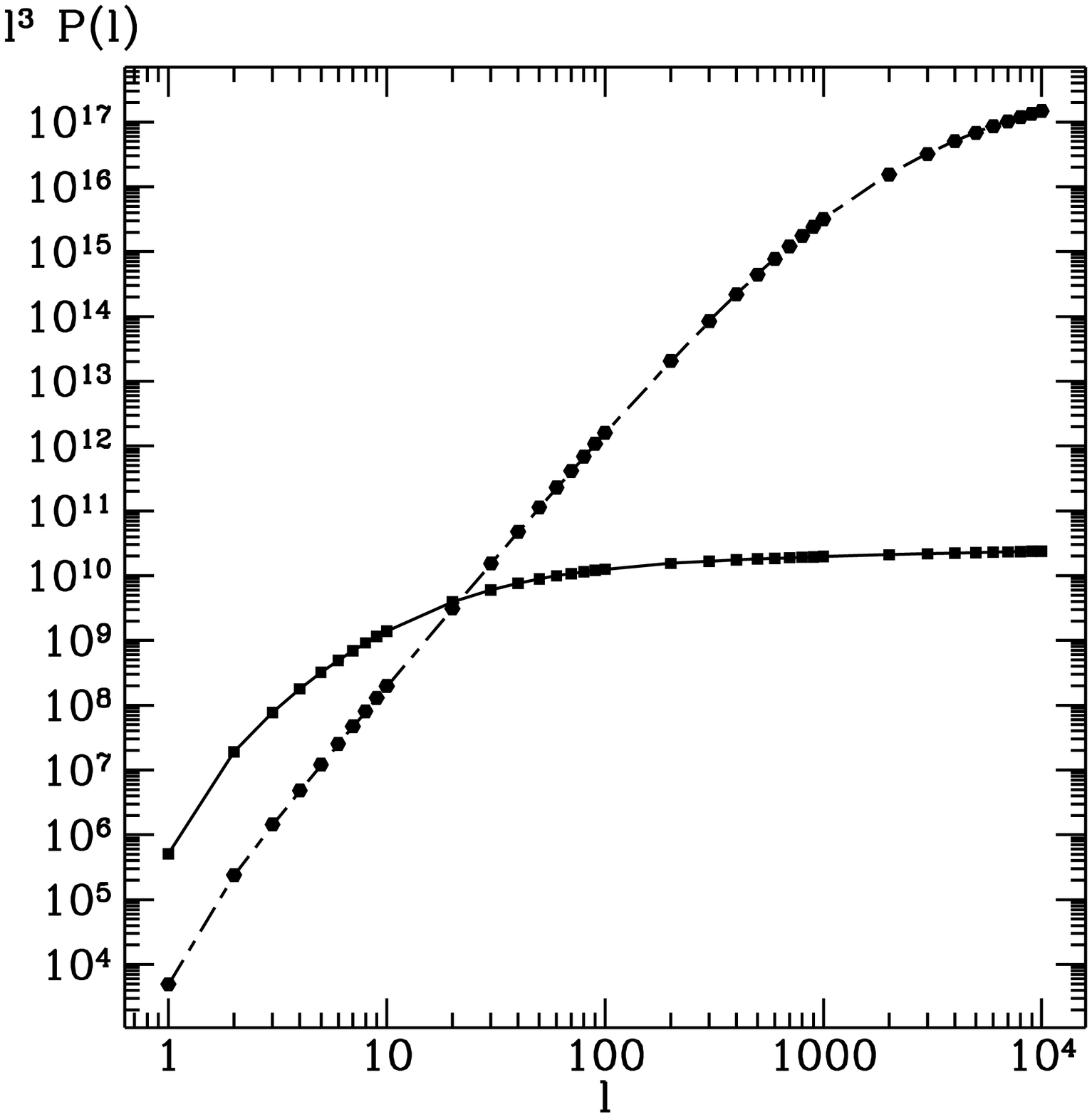,width=7.5cm}
             }
        \end{flushleft}
\caption{\label{fig5}
The power spectrum $P(l)$ and the induced mass fluctuation $l^{3} P(l)$
as a function of $l$ for matter density perturbations in the composite model
(in arbitrary units). The lines connecting the points $l\in {\bf N}$ are
shown for clarity. (${\bf A}$) $\&$ (${\bf B}$): A purely radiation
dominated cycle, $\alpha
\ll 1$, $a_{max}\ll a_{eq}$.
(${\bf C}$) $\&$ (${\bf D}$): A cycle which just
reaches the matter dominated epoch, $\al=1 ,\  a_{max}=a_{eq}$.
(${\bf E}$) $\&$ (${\bf F}$):
A cycle including a short matter dominated epoch with
$\alpha=4$ (square points, maximum of $P(l)$ in $({\bf E})$ at $l\sim 4$) and
a cycle including a long matter dominated epoch ($a_{max}\gg a_{eq}$)
with $\alpha=1000$ (hexagonal points, maximum of $P(l)$ in $({\bf E})$
at $l\sim 1000$).}
\end{figure}

This leads to the following behavior: if a cycle is purely 
radiation dominated  (i.e. $a< a_{eq}$),
the power spectrum takes its maximum for the smallest value of $l$.
If a sequence of cycles approaches and finally enters  
the matter dominated era, then there will be a 'critical cycle' from 
which on the maximum of $P(l)$ is no longer the largest scale, $l=1$
but a scale comparable to $l_{eq}>1$. 

\begin{figure}
        \begin{flushleft}
        \mbox{
                (A)\psfig{file=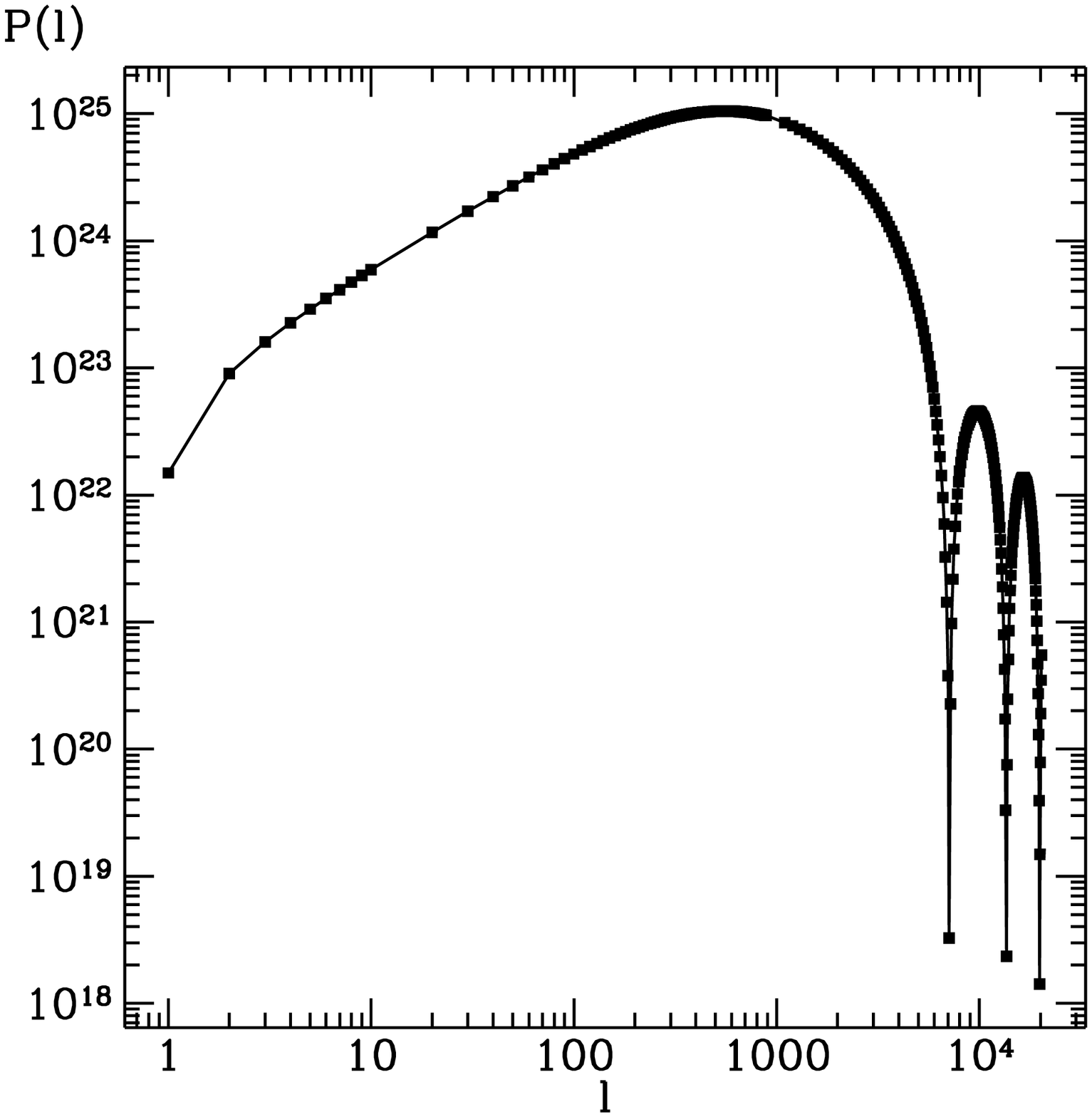,width=7.5cm}
                (B)\psfig{file=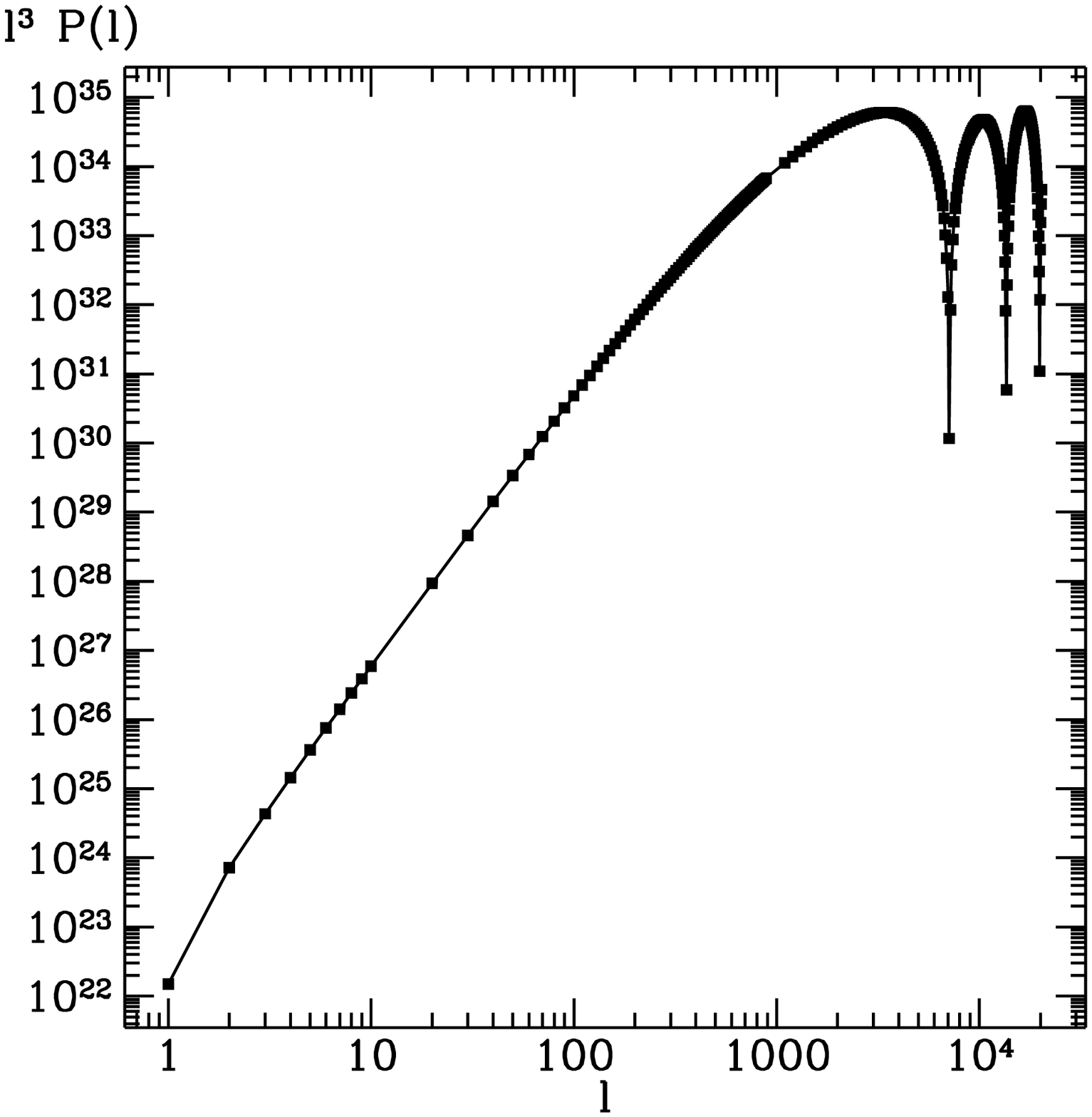,width=7.5cm}
             }
        \end{flushleft}
\caption{\label{fig6}
(${\bf A}$): The power spectrum $P(l)$ in the composite model
as a function of $l$, when matter
becomes relativistic close to the crunch (arbitrary units). The cycle includes
a long matter dominated epoch $\al=1000$ (The solid line simply connects the
evaluated points).
(${\bf B}$): The induced mass fluctuation $l^{3} P(l)$
for the same cycle as in (${\bf A}$).}
\end{figure}

Obviously equation
(\ref{comp2}) is independent of $l$ and hence  the shape of the
power spectrum is exclusively determined by the scale invariance condition at
horizon crossing and does not change during the subsequent growth of
fluctuations. This is not the case, when the particles become
relativistic. Then the evolution equation (\ref{comp1}) does indeed
depend on $l$ and the shape of the power spectrum changes when the
cycle approaches the crunch: further local maxima will occur due to the 
oscillating  behavior of solutions of (\ref{comp1}), 
but the global maximum of the power spectrum remains the same. 
Numerical solutions for this last case are shown in Fig.~\ref{fig6}.

\begin{figure}
        \begin{center}
        \mbox{
                \psfig{file=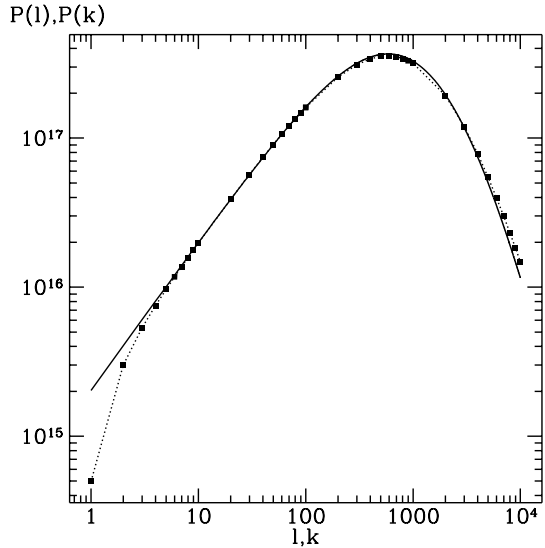,width=7.5cm}
             }
        \end{center}
\caption{\label{fig7}
The discrete power spectrum $P(l)$ for the composite model with $\al=1000$
(square points connected by dots) in comparison with the continuous flat space
analogon $P(k)$ (solid line).}
\end{figure}

In a spatially flat universe containing matter and radiation, the power 
spectrum for matter density perturbations can be approximated at 
times $t\gg t_{eq}$ by
	\be\label{p9}
	P(k,t)\cong \frac{C^{2} k t^{4}}{(1+(k/k_{eq})^{2})^{2}},
	\ee
which is similar to (\ref{power4}), only that $k$ is continuous and the
additional decrease of $P$ for small values of $k$
does not occur (In an open universe the power spectrum even starts to 
increase for the largest scales). 
Therefore a
measurement of the power spectrum at very large scales (even before 
a cycle has reached it's maximum
expansion) would (in principle) be a way to decide, whether our 
universe is flat, open or closed. To see the departure of (\ref{power4}) 
from (\ref{p9}) for
large scales, we have plotted both curves together in Fig.~\ref{fig7}.
A similar behavior has also been found in \cite{WS}.

\subsection{Interpretation of the Results, Entropy Production in the
Previous Cycle}
We have thus found that a short time before the big crunch, the mass 
fluctuation, $\De^2(l,t)=l^3P(l,t)$ is scale independent in a pure 
radiation universe and decreases towards large scales, $l\le l_{eq}$ in a 
matter/radiation universe.

Furthermore,  $\De^2(l,t)$ diverges at the big crunch. Therefore, at
least briefly before the big crunch, linear perturbation theory is no
longer applicable. In the pure radiation case, we expect non linear 
effects to stop gravitational instability and prevent black hole
formation at least on small scales. The production of gravitational
entropy is thus probably not very significant.

However, if the universe undergoes an intermediate matter dominated
period,  $\De^2(l,t)$ tends to raise towards smaller scales, approaching
a very mild, logarithmic growth for $l > l_{eq}$ (see Fig.~\ref{fig5}(D)). We also 
know from the 
corresponding flat universe analysis, if fluctuations grow non--linear 
before, due to contraction, the universe becomes radiation dominated 
again, non--linear gravity and the log--raise towards small scales will 
lead to the collapse of small scales and probably to the 
formation of small black holes.

If we want to prevent excessive black hole and entropy formation
in the cycle previous to the present one, we thus have to require that 
perturbations never get strongly nonlinear. This yields a limit for the
radiation entropy in the previous cycle. To illustrate this, let us 
assume  that in the present cycle perturbations get non--linear,  
$\De^2(l,t) \approx 1$ around a redshift of $z\sim 10$, $T\sim 30K$. For
this not to happen in the previous cycle, we have thus to require
$T_{\min}>30K$ or $S= (\la_2/T_{\min})^3 < 10^{84}$. 

The radiation entropy of the previous cycle thus has to be at least a 
factor of $10^3$ times smaller than the present entropy. We therefore
require that most of the radiation entropy of the universe at present,
$S\ge 10^{87}$, was produced in the form of gravitational entropy from
small fluctuations during the previous cycle. Unfortunately, we do not
have a quantitative description for the entropy of the gravitational
field (except in the extreme cases of linear perturbations\cite{BM} or 
black holes), but it is certainly related
to the clumpiness of the matter which is determined by the Weyl part
of the curvature \cite{Pe}.

We now postulate, that during the quantum gravity epoch of big 
crunch/big bang passage
 the entropy of the gravitational field is completely transformed
into radiation entropy and the new cycle starts out from a state with 
vanishing gravitational entropy, a homogeneous and isotropic Friedmann 
Lema\^{\i}tre universe. At first this postulate might seem somewhat ad
hoc, but it is actually just what happens if a black hole evaporates
due to Hawking radiation.

It is thus feasible that most of the entropy production in the
 universe is actually due to mild gravitational 
clustering in the previous cycle and not due to local non--thermal
processes. In this way it can be achieved that the radiation entropy in 
the previous cycle was much (several orders of magnitude) lower than in
the present cycle.

\section{Conclusions}
We have revisited  the oscillating universe and shown how it can yield 
a coherent solution to the flatness or entropy and the horizon 
problems of standard cosmology. 
We have analized linear gravitational perturbations in a closed 
universe consisting of matter and radiation. We can set an upper limit
on the radiation entropy of the previous cycle which is at least a factor
$10^3$ below the entropy of the present universe. We thus postulate that
most of the radiation entropy in the present cycle was produced as 
gravitational entropy by linear or mildly non--linear gravitational
clustering in the previous cycle. During the quantum gravity big 
crunch/ big bang era, this gravitational entropy must then be 
transformed into radiation entropy. 

Due to the lack of of a theory of quantum gravity, we have no
precise idea how this is accomplished. Nevertheless, this is exactly 
what happens if black holes evaporate!

The reader may now object that we postulate the emergence of a 
Friedmann Lema\^{\i}tre universe out of the quantum gravity era, whereas 
homogeneity and isotropy are naturally obtained in some inflationary 
models, e.g. chaotic inflation. However also in chaotic inflation, 
where homogeneity and isotropy is achieved by blowing up small scales,
one has to require a cut--off of fluctuations at some very small scale,
typically around Planck scale.

We consider it to some extend a matter of taste which of the two 
requirements for quantum gravity is more 'restrictive'; that it leads
to a cut--off of fluctuations around the Planck scale or that it leads
to the transformation of gravitational entropy into radiation entropy
in very high curvature regions. Nevertheless, it is a weakness of our model,
that we cannot propose a clear picture of how this transformation might 
take place. We plan to address this problem in the future.

In our approach the creation of initial density fluctuations is not 
discussed. They might be created in (or left over from)
the Planck era or they might build up during a phase transition in the
early universe (topological defects) or by any other scale invariant
process, like the self ordering of a global scalar field on Hubble
scale.

Also the monopole problem present in simple GUT's is not addressed. It may be
solved by breaking electromagnetism at high temperature (see \cite{LP}),
or it may just not exist if the GUT idea of a simple unifying gauge group
at high energies is not realized.

Let us summarize the predictions and limitations 
of the oscillating universe in the following list mentioning the 
main  problems of standard cosmology:
\begin{itemize}
\item {\em Flatness, age, entropy problem:} Is solved, but predicts 
	$\Om=1+\ep$.
\item {\em Horizon problem:} Is solved, but a method to calculate the true age
	of the universe is still missing.
\item {\em Cosmological constant problem:} Is not addressed.
\item {\em Monopole problem:} Is not addressed, may be solved along the lines
	mentioned above.
\item {\em Initial fluctuations:} Are not addressed. May be left over from 
	a quantum era or may be due to topological defects.
\item {\em Fine tuning:} There seems not to exist a serious fine
	tuning problem.
But since the entropy was so much smaller in the previous cycle, the present
cycle is the first one in which galaxies, stars and human beings can form.
\end{itemize}
Clearly, the easiest way to falsify this model would be to measure
$\Om<1$. Many measurements of mass to light ratios in galaxies and 
clusters hint that the density parameter is indeed less than 1. However, 
these are just measurements of clustered matter. An {\em upper} limit on 
matter which is not or only weakly clustered on these scales is much more 
uncertain. Furthermore, if $\Om\equiv 1$ we will never be able to decide
whether $\Om=1+\ep$ or $\Om=1-\ep$, and  other means to distinguish 
this scenario from, e.g., inflationary models have to be developed.

\vspace{1cm}

\noindent {\large \bf Acknowledgement}\\
We thank Norbert Straumann and Mairi Sakellariadou for stimulating discussions.
This work is supported by the Swiss national science foundation.

\newpage

\noindent
{\LARGE \bf Appendix}
\vspace{1cm}\\
\appendix
\section{Scalar Harmonic Functions}\label{A1}
In the closed universe
scalar quantities like $D$ can be expanded in a complete set of scalar 
harmonic functions ${\cal Y}_{{\bf k}}({\bf x})={\cal Y}_{{\bf k}}
(\chi,\theta,\phi)$ on the three sphere ${\bf S}^{3}$: 
	\be\label{expand}
	D(x)=\sum_{{\bf k}}{\cal Y}_{{\bf k}}(\chi,\theta,\phi)
	D_{{\bf k}}(t),
	\ee
where ${\bf k}=(l,j,m)$, $l=0,1,2,3,\ldots$, $j=0,1,\ldots ,l-1$, $m=-j,-j+1,
\ldots,j$. The variables $\chi\in [0,\pi]$, $\theta\in [0,\pi]$ and 
$\phi\in [0,2\pi]$ denote the angles describing the position on the three
sphere. The functions ${\cal Y}_{{\bf k}}$ satisfy the 
Laplace-Beltrami equation with eigenvalue $-k^{2}$:
	\[(\Delta +k^{2}){\cal Y}_{{\bf k}}=0.\]
Here $\Delta\equiv \nabla^{j}\nabla_{j}$ denotes the three-dimensional 
Laplacian on ${\bf S}^{3}$,  
$k^{2}=l(l+2)K$ and for our case of interest $K>0$ (In most of  the sequel we 
set $K=1$). The harmonic 
functions ${\cal Y}_{{\bf k}}({\bf x})$ are given by
	\[{\cal Y}_{{\bf k}}({\bf x})=\Pi_{\beta j}^{(+)}(\chi) 
	Y_{jm}(\theta,\phi),\qquad \beta^{2}=k^{2}+K=(l+1)^{2}K,\]
where $Y_{jm}(\theta,\phi)$ are the usual spherical harmonics on ${\bf S}^{2}$
and the $\Pi_{\beta j}^{(+)}$ can be expressed in terms of generating functions
	\[\Pi_{\beta j}^{(+)}(\chi)=i^{j}\frac{\sin^{j}\chi}{(M^{j}_{\beta})
	^{1/2}}\left(
	\frac{d}{d\cos\chi}\right)^{j+1}\cos(\beta\chi),\]
where $M^{j}_{\beta}$ is the normalization factor
	\[M^{j}_{\beta}=(\pi/2)\prod_{n=0}^{j}(\beta^{2}-n^{2}).\]
The normalization of the functions ${\cal Y}_{{\bf k}}({\bf x})$ 
is as usual
	\[\int_{{\bf S}^{3}}d^{3}x\: h^{1/2}({\bf x}) {\cal Y}_{{\bf k}}
	({\bf x})
	{\cal Y}_{{\bf k}'}^{*}({\bf x})=\de_{{\bf k},{\bf k}'},\]
where $h({\bf  x})$ is the determinant of the 3-metric of constant curvature
$K=1$ and $\de_{{\bf k},{\bf k}'}$ is the Kronecker delta
	\[\de_{{\bf k},{\bf k}'}=\left\{\begin{array}{ll}
	1,\quad & \mbox{if}\quad {\bf k}\equiv (l,j,m)={\bf k}'
	\equiv (l',j',m') \\
	0, & \mbox{else}.\end{array} \right.\]
Furthermore we choose the phases of ${\cal Y}_{{\bf k}}$ such that 
${\cal Y}_{{\bf k}}^{*}({\bf x})={\cal Y}_{-{\bf k}}({\bf x})$,
with $-{\bf k}\equiv (l,j,-m)$.
(See e.g. \cite{dolg} or \cite{pafu} for further details).

\section{The Passage between two Cycles}
In this Appendix we want to show briefly, how the transition from a 
big crunch to a subsequent big bang can be described by an effective model.
The main idea is the appearance of a signature change in the metric from
 Lorentzian to Euclidian and back. By this mechanism, the singular
behavior of spacetime at $a=0$ disappears and the topology of the transition 
region is that of ${\bf S}^{4}$. 

In analogy to the change of signature idea of Hartle \& Hawking 
\cite{haha} in quantum cosmology, Ellis \cite{elli} and 
Ellis et.~al.~\cite{elal}  have shown that the classical
Einstein field equations allow a change of signature when the metric is
allowed to possess a mild singularity. The classical case leads to interesting 
possibilities for the description of an oscillating universe. 

The signature 
change is implemented into the metric by introduction of a lapse function
$n(\tau)$:
	\bea\label{fri5}
	ds^{2}&=&-n(\tau)d\tau^{2}+a^{2}(\tau)\left[\frac{dr^{2}}{1-kr^{2}}
	+r^{2}(d\theta^{2}+\sin^{2}\theta \:d\vi^{2})\right].
	\eea
Here $\tau$ denotes cosmic time: $d\tau=a\:dt$.
For the discontinuous choice of the lapse function $n(\tau)=\ep$ with
$\ep=\pm 1$, there exists a surface of change $\Si$, where the metric
changes its signature. From (\ref{fri5}) and Einstein's equations one 
derives the Friedmann and Raychaudhuri equations for the scale factor 
$a(\tau)$, which then hold in the regions $V_{+}$, where $\ep=+1$ and
in $V_{-}$, where $\ep=-1$, but not on the 
surface of signature change $\Si$, since there the metric tensor is not 
invertible.
By choosing suitable (physically motivated) jump conditions 
on  $\Si$, one can find solutions for the scale factor which pass
continuously through the surface of signature change \cite{elli,elal}. 

For the simple case of a scalar field $\phi\in {\bf R}$ with Lagrangian
	\[{\cal L}=\frac{1}{2}\partial_{\mu}\phi\partial^{\mu}\phi-V(\phi),\] 
such that $\dot{\phi}=0$  (the no rolling case), Ellis et.~al.~have shown 
that the  scale factor behaves  for $k=1$ like
	\be\label{scal}
	a(t)=\left\{\begin{array}{ll}
		H^{-1}\cos(H\tau),\qquad &-\pi/(2H)\leq \tau\leq 0 \\
		H^{-1}\cosh(H\tau), &\tau\geq 0 \; . \end{array}\right.
	\ee
The corresponding space-time has no boundary and is 
geodesically complete (i.e. it has no singularity and the geodesics can be
continued smoothly through the surface of
signature change). Only the length of the tangent vector jumps
  for photons and spacelike geodesics at the surface of signature 
change $\Sigma$. Obviously the scale factor given by (\ref{scal}) inflates
for $\tau>0$. We do not have an equivalent simple example with the same 
nice features which does not inflate. However 
there are other (rolling) solutions to get a 
successful exit from inflation, 
but these solutions do not have the 'no-boundary' property of the above
mentioned case. For a detailed discussion see \cite{elli} 
and \cite{elal}. Our main point is, that there is a
possibility to continue the evolution of the universe smoothly through the 
crunch which might serve as an effective theory for the passage between two 
subsequent cycles.

\subsection{Geometric Representation}
It is interesting to note that only geodesics describing massive
particles at rest
pass through the point where $a=0$, whereas a particle with velocity $v_{\Si}
\not=0$ on $\Si$ enters the Euclidian regime with an angle $\al$ as shown 
in Fig.~\ref{fig8} (The coordinates shown are $t$ and $r$. The angles 
$\theta$ and $\phi$ are suppressed, since we assume them to be constant 
for the indicated geodesic).
\begin{figure} 
\caption{\label{fig8}
A Geodesic of ${\bf S}^{2}$ with initial velocity $v_{\Sigma}>0$.}
\end{figure}

The coordinates of the plane with angle $\al$ in the $y-z$ plane, are 
	\[(x,y,z)=(x,-z\tan\al,z).\]
With polar coordinates chosen as
	\[x=\cos\theta\cos\vi,\quad y=\cos\theta \sin\vi,\quad z=-\sin\theta,\]
where $\vi\in[0,\pi]$ and $\theta \in[0,\pi/2-\al]$, one finds 
	\[\cos\theta\sin\vi=\sin\theta\tan\al.\]
Hence a geodesic on ${\bf S}^{2}$ through $\vi=\theta=0$ is given by
	\[\vi(\theta,\al)=\arcsin[\tan\theta\cdot \tan\al].\]
Here $\theta=H \tau$ is the time coordinate. The velocity $v_{\Si}$ on 
$\Si$ then is
	\[v_{\Si}\equiv\frac{d\vi}{d\theta}|_{\theta=0}=\tan\al,\]
indicating that a point particle which is not at rest, does not reach the point
where $a=0$. A photon with $v_{\Si}=1$ enters with the angle $\al=\arctan 1$
which is $\pi/4$. Hence a photon is not represented by the
boundary at $\alpha=\pi/2$, but by a line inside the Euclidian regime.

\end{document}